\documentclass[a4paper,10pt]{article}
\usepackage{multicol}
\usepackage{amsmath}
\usepackage{amssymb}
\usepackage{graphicx}
\usepackage{float}
\usepackage{amsmath}
\usepackage{bm}
\usepackage{amsfonts}
\usepackage{balance}
\usepackage{titling}
\usepackage{cite}
\usepackage{color}
\makeatletter
\newcommand\figcaption{\def\@captype{figure}\caption}
\makeatother
\newcommand{\ct}[1]{{\textsuperscript{{\cite{#1}}}}}
\newcommand{\bee}{\begin{equation}}
\newcommand{\ee}{\end{equation}}
\newcommand{\beea}{\begin{eqnarray}}
\newcommand{\eea}{\end{eqnarray}}

\newcommand{\vv}[1]{{\mathbf #1}}

\newcommand{\ci}{\textrm{i}}
\newcommand{\corr}[1]{ #1}
\begin{document}
\title{Axion-plasmon-polariton hybridization in a graphene periodic structure}
\author{ {Daqing Liu$^{1}$
, \, Xiuqin Hua$^{1}$, Dong Sun$^1$, Xingfang Jiang$^1$ and Xiang Zhao$^2$} \\
{\small \it $^{1}$ School of Microelectronics and Control Engineering, Changzhou University, Changzhou, 213164, China}\\
{ \small \it $^2$ School of Science, Xi'an Jiaotong University, Xi'an, 710049, China }\\
 }
\date{}
\maketitle

 \begin{abstract}
This study investigated the hybridization between an axion and plasmon polariton, attributed to the coupling achieved by combining modified electrodynamics and hydrodynamic approaches on a plasmon-polariton in a graphene periodic structure. The enhancement of the effective coupling was also studied. Furthermore, corrections for the axion and the lowest plasmon -polariton hybridization state spectra have been presented. An observable was  proposed to detect axions, which is significant even when the effective coupling is not large, especially as the axion mass decreases. The study shows that the resulting structure provides a sensitive and wide-mass-spectrum platform for detecting axions at the sub-meV scale.
 \end{abstract}



\section{Introduction}

An axion was originally postulated to address the strong charge-parity problem\ct{1,2}. It is considered that this hypothetical particle  has an extremely small mass (possibly smaller than meV, for instance, in the range of $10^{-3} - 1$ meV) and couples very weakly with quarks, leptons, and photons (the coupling between the axion and common matters was implied as $g\leq 10^{-10}\text{GeV}^{-1}$). This particle was considered to be one of the prime dark matter candidates\ct{3,4,5,6,7,8,9}. Recently, dark-matter axion detection emerged as a promising research field\ct{10,11,12,13,14,15,16,17,18,19}, and many studies have used a variety of techniques to detect axions\ct{20,21,xmass,quax,lat,madmax,26,haloscope,jcap,plb}.

Several studies\ct{16,17,19} have used the interactions between axions and plasma to detect axions using photoelectric techniques\ct{9,20,haloscope}. However, when using such an approach, it is challenging to control the parameters. \corr{For instance, refs. \cite{16,17} take advantage of laser-plasma experiment to observe axion signature. However, the need for the strong magnetic field and high electric density makes that one can only observe the signature in the interior of neutron stars. Not to mention how to constrain the plasma and how to control the plasma temperature in a laboratory.} In this study, we propose a novel scheme to observe the axion effect in the laboratory. In the presence of strong background magnetic fields, the coupling between an axion and plasmon polariton (PP), which is not plasma but a collective excited state in a material, occurs in a graphene periodic structure. We show that the effective coupling between an axion and PP can be enhanced by adjusting certain parameters, such as Fermi energy, the background magnetic field, and graphene interlayer distances. Corrections to the plasmon-polariton/axion hybridization state spectra were presented and an observable to detect axion was suggested subsequently. The signal is significant even when the effective coupling is not large, especially as the axion mass decreases. Furthermore, as the significance does not rely on the phenomenon of resonance, there is a broad detectable range of axion mass spectrum using the observable. \corr{At last, parameters we used here, such as external magnetic field, layer distance, and carrier density (controlled by gate voltage) are easier to be tuned in the laboratory since carriers are be confined in the graphene materials.}
The proposed method therefore provides an effective platform for detecting axion particles.


\section{Modified electromagnetism in a graphene structure}
A three-dimensional periodic structure of graphene layers embedded in a medium with permittivity $\epsilon$ (we set $\epsilon$ to be equal to vacuum permittivity $\epsilon_0$) is shown in Fig. 1. The graphene layers parallel to
the plane defined by $z=0$ are assumed to be infinite for simplification. The distance between the layers is represented as $d$. \corr{Considering the graphene carrier as a 2D electron gas in graphene layer, we suppose that each} graphene layer is $N$-doped and has an equal Fermi energy, $E_F>0$, with $n_0=\frac{k_F^2}{\pi}=\frac{E_F^2}{\pi \hbar^2 v_F^2}$ where $\hbar k_F$, $v_F$, and $n_0$ are Fermi momentum, Fermi velocity($1\times 10^6\text{m/s}$) and two-dimensional(2D) carrier equilibrium density, respectively. The hydrodynamic equation of the carrier is expressed as follows\ct{prb96}
\bee
m_g n\frac{\partial \vv{v}}{\partial t}+m_g n(\vv{v}\cdot\nabla)\vv{v}=-en(\vv{E}+\vv{v}\times \vv{B})-\nabla P,
\ee
\corr{where $e$, $n$, $\vv{v}$, $\vv{E}$ and $\vv{B}$ are carrier charge, carrier density, fluid velocity, electric field, and magnetic field, respectively. }In the above equation $m_g=\hbar k_F/v_F$ is the effective mass of carriers at the Fermi surface\ct{prb96,mg}, which is at the order of $0.1 - 1\text{eV}$ and can be tuned by $E_F$, and $P=\frac{\hbar v_F}{3\pi}(\pi n)^{3/2}$ ($\nabla P=\frac{\hbar v_F}{2}\sqrt{\pi n}\nabla n$) is the carrier pressure in the graphene. \corr{Note we here assume that the pressure has the same functional form as that of the 2D gas.}

\begin{center}
\begin{minipage}{0.85\textwidth}
\centering
\includegraphics[width=2.6in]{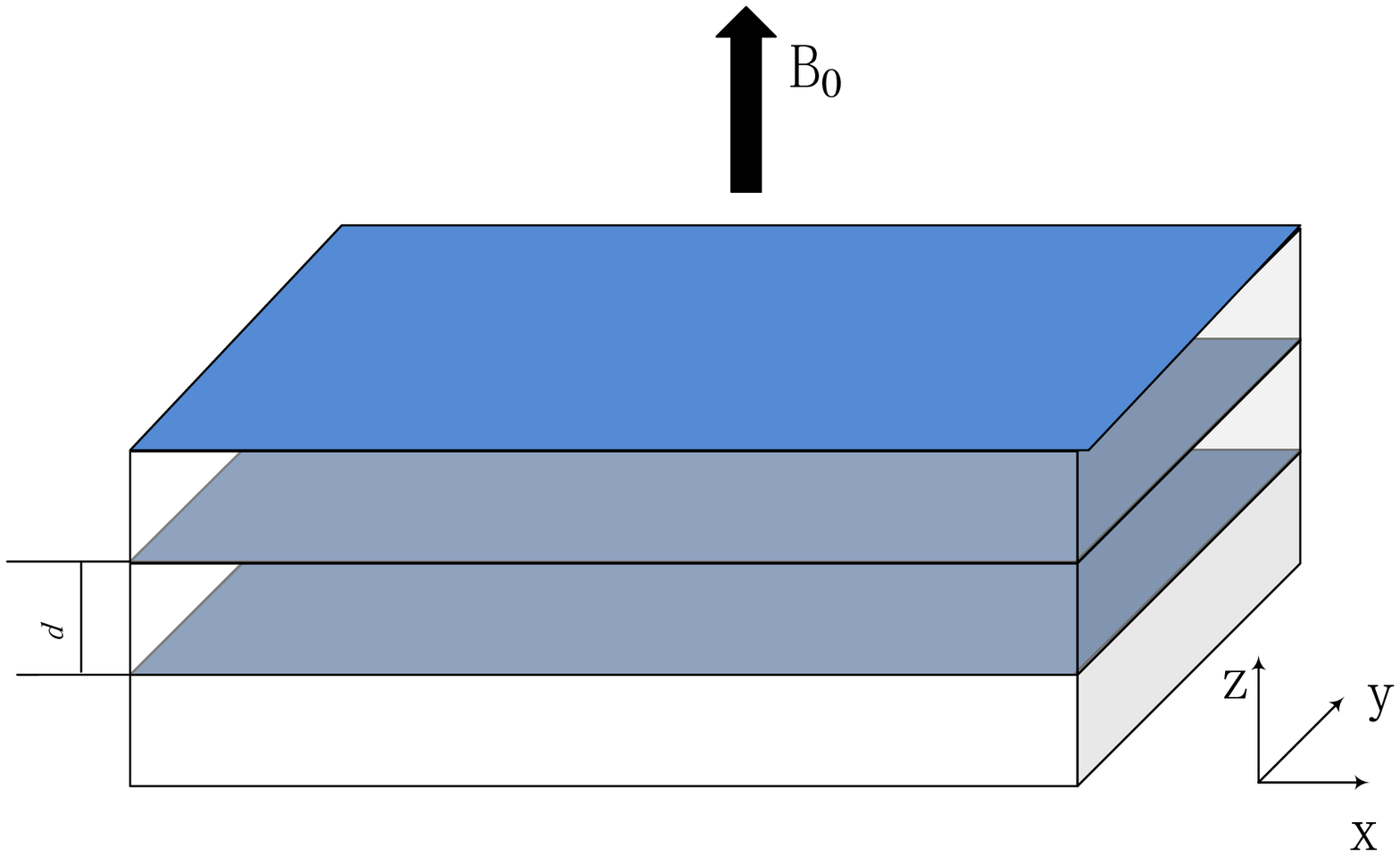}%
\figcaption{Three-dimensional periodic structure of graphene ribbons embedded in the medium.
}
\end{minipage}
\setlength{\intextsep}{0.in plus 0in minus 0.1in} 
\end{center}

The continuity equation  is
 \bee
 \frac{\partial n}{\partial t}+\nabla\cdot(n\vv{v})=0.
 \ee

When $d$ is small enough, the modified Maxwell equations become:
\bee
\left\{
\begin{array}{l}
  \nabla\cdot \vv{E}=-e\rho/\epsilon_0+cg\nabla\phi\cdot\vv{B}, \\
  \nabla\times \vv{B}=\frac{\partial\vv{E}}{c^2\partial t}+\mu \vv{j}/d+\frac{g}{c}(\vv{E}\times \nabla \phi-\vv{B}\frac{\partial\phi}{\partial t}), \\
  \nabla\cdot \vv{B}=0, \\
  \nabla\times \vv{E}=-\frac{\partial \vv{B}}{\partial t},
\end{array}
\right.
\ee
where $\rho=(n-n_i)/d=(n-n_0)/d$ is the three-dimensional(3D) particle density; $\mu$ is the vacuum permeability; $\vv{j}$ is the linear current density; $\phi$ is the axion field; and $g$ is the coupling between the axion and electromagnetic fields with the dimension $[M^{-1/2}T^{1/2}]$. Here, we assume that $d$ is sufficiently small such that the boundary conditions at each graphene layer can be ignored.

The Klein-Gordon equation for the axion field is:
\bee
\frac{\partial^2\phi}{c^2\partial^2 t^2}-\nabla^2\phi+\frac{m_\phi^2 c^2}{\hbar^2} \phi
=-g\epsilon_0 \vv{E}\cdot\vv{B},
\ee
where $m_\phi$ is the axion mass.

Subsequently, we cast a strong homogeneous background magnetic field, $B_0$, perpendicular to graphene layers. To study the system behavior, we linearize the above equations.
\corr{In order to do it, we first write $n=n_1+n_0$, where $n_1$ is the density perturbations around the carrier equilibrium density $n_0$. Here we assume that the system is far away from the neutral point and the perturbation is not large, that is, $|n_1|\ll n_0$. With the assumption we have:}

\beea
&&\frac{\partial \vv{v}}{\partial t} = -\frac{e}{m_g}(\vv{E}_{\parallel}+\vv{v}\times \vv{B}_0)-\frac{\pi \hbar^2}{2m_g^2}\nabla n_1,\label{ln_heq}\\
&&\frac{\partial n_1}{\partial t}+n_0\nabla\cdot\vv{v} =0,\\
&&\left\{
\begin{array}{l}
  \nabla\cdot \vv{E}=-e n_1/(d\epsilon_0)+cg\nabla\phi\cdot\vv{B}_0, \\
  \nabla\times \vv{B}=\frac{\partial\vv{E}}{c^2\partial t}+\mu \vv{j}/d -\frac{g}{c}\vv{B}_0\frac{\partial\phi}{\partial t}, \\
  \nabla\cdot \vv{B}=0, \\
  \nabla\times \vv{E}=-\frac{\partial \vv{B}}{\partial t},
\end{array}
\right. \label{ln_maxwell}\\
&&\frac{\partial^2\phi}{c^2\partial^2 t^2}-\nabla^2\phi+\frac{m_\phi^2 c^2}{\hbar^2} \phi
=-g\epsilon_0 \vv{E}\cdot\vv{B}_0, \label{ln_kg}
\eea
where $n_1=n-n_0$. In the above equations $\vv{v}$ and $\vv{j}$ have only $x$ and $y$ components, and $\vv{E}_{\parallel}$ is the projection of $\vv{E}$ in the xy-plane. In physics, the electrostatic perturbation along the background magnetic field will produce the axion field through Eq. (\ref{ln_kg}). Then, the axion field, acting as a source, will generate the electromagnetic field according the modified Maxwell equation (Eq. (\ref{ln_maxwell})). The electromagnetic field will drive the carrier oscillation, the plasmon polariton, in graphene layers according to Eq. (\ref{ln_heq}). \corr{The above process is reversible. Therefore, axion and plasmon polariton hybridize and axion-plasmon-polariton hybridization states exist in the structure. If we symbol the angular frequency of plasmon polariton as $\omega_{pp}$,  the hybridization does not require that $m_\phi c^2\simeq \hbar \omega_{pp}$ provided $m_\phi c^2,\hbar \omega_{pp} \ll \hbar\omega_p$, where $\omega_p$ is the plasmon frequency. Note that the condition $m_\phi c^2\simeq \hbar \omega_{pp}$ is necessary if one takes advantage of the resonance of the axion and the plasmon polariton. }

We assume that the hybridization states propagate in the $xz$ plane with the wave number $\vv{q}=(q\sin\theta,0,q\cos\theta)$, and angular frequency $\omega$, where $\theta$ is the angle between the direction of the axion propagation and the background magnetic field. \corr{Moreover, we also assume that the quantities, $n_1$, $\vv{v}$, $\phi$, $\vv{E}$ and $\vv{B}$ go as $e^{\ci\vv{q}\cdot\vv{r}-\ci \omega t}$}. The modified Helmholtz equations in the system transform into:
\bee
\kappa^2\vv{B}=-\frac{g\omega}{c}\phi\vv{q}\times\vv{B}_0+ \ci \mu \vv{q}\times\vv{j}/d,
\ee
for the magnetic field, and
\bee
\kappa^2\vv{E}=\ci\frac{e n_1}{d\epsilon_0}\vv{q} +cg\phi(\vv{q}\cdot\vv{B}_0)\vv{q}+\ci \omega\mu\vv{j}/d-\frac{g\omega^2}{c}\phi\vv{B}_0,
\ee
for the electric field, where $\kappa^2=q^2-\frac{\omega^2}{c^2}$.

Decomposing all the fields into transverse and longitudinal parts, for instance, $\vv{E}_L=\hat{q}\hat{q}\cdot\vv{E}=E_L \hat{q}$ and $\vv{E}_T=\vv{E}-\vv{E}_L$, \corr{with $\hat{q}$ standing for the unit vector in the direction of $\vv{q}$}, we have
\beea \label{decom}
\kappa^2 D_L &=& \ci\omega\mu j_L/d+\ci\frac{eqn_1}{d\epsilon_0},\notag \\
\kappa^2 \vv{D}_T &=& \ci\omega\mu \vv{j}_T/d-cgq^2\phi\vv{B}_{0T},
\eea
where $\vv{D}=\vv{E}-cg\phi \vv{B}_0$.

\corr{It seems that the longitudinal mode decouples with the axion field, as shown in the first equation in Eq. (\ref{decom}). However, since the linear current density has $x$ and $y$ components only, it usually has both longitudinal and transverse parts, unless the hybridization states propagate along the $z$-axis. Therefore, just as the transverse component behaves, the longitudinal component will also couple with the axion field. This is one of the crucial differences between the graphene structure and ordinary plasma.}

\section{Dispersion of axion-plasmon-polariton hybridization state}
To obtain the dispersion of the hybridization state, $\vv{j}, \vv{v}$, and $\vv{E}$ should be eliminated from the above equations. For linear current density, we ignore the interband contribution and write the equation as
\bee
\vv{j}=-en_0 \vv{v}.
\ee
We finally obtained:
\beea \label{res1}
\frac{\ci g m_g \omega_c q \cos\theta}{d\kappa^2}n_1+[\kappa^2+\frac{\omega_\phi^2 }{c^2}+\frac{g^2 m_g^2\omega_c^2 c\epsilon_0}{e^2\kappa^2} (q^2\cos^2\theta-\frac{\omega^2}{c^2})]\phi &=& 0 ,\notag \\
\frac{ \ci g d m_g c^3 q^3 \omega_p^2 \omega_c\epsilon_0 h_1\cos\theta \sin^2\theta}{e^2 f_1}\phi+
[1-\frac{c^2 q^2 (2\omega_p^2+v_F^2\kappa^2)h_1\sin^2\theta}{2f_1}]n_1 &=& 0
\eea
where $h_1=\omega_p^2+c^2\kappa^2$,
$f_1=\omega^2 h_1^2-c^4\omega_c^2\kappa^4$, $\omega_\phi=\frac{m_\phi c^2}{\hbar}$, $\omega_c=\frac{eB_0}{m_g}$ is the carrier cyclotron frequency at the Fermi surface, and $\omega_p^2=\frac{e^2 n_0}{d\epsilon_0 m_g}$ is the plasmon frequency related to carrier density $n_0/d$ and carrier effective mass $m_g$.

To understand the physics behind the above equations, we first set $g=0$, that is, no coupling exists between the axion and electromagnetic fields. The first equation in (\ref{res1}) shows the free axion field dispersion, $\kappa^2+\frac{m_\phi^2 c^2}{\hbar^2}=0$, as expected. The second equation in (\ref{res1}) presents the results of PP dispersion, given by
\bee \label{res2}
1-\frac{q_0^2(1+q_0^2-\omega_0^2)\sin^2\theta(2+v_{F0}^2(q_0^2-\omega_0^2))}
{2[\omega_0^2(1+q_0^2-\omega_0^2)^2-\omega_{c0}^2(q_0^2-\omega_0^2)^2]}
=0,
\ee
where the following dimensionless quantities have been introduced, $\omega_0=\omega/\omega_p$, $\omega_{c0}=\omega_c/\omega_p$, $v_{F0}=v_F/c$, and $q_0=cq/\omega_p$.

If we turn off the background magnetic field, that is, $\omega_{c0}=0$, the two solutions can be expressed as:
\beea
(\omega/\omega_p)^2
&=& \frac{1}{4}[2+2q_0^2+q_0^2 v_{F0}^2\sin^2\theta \pm \sqrt{(2+2 q_0^2+q_0^2 v_{F0}^2\sin^2\theta)^2-8 q_0^2(2+q_0^2 v_{F0}^2)\sin^2\theta} ] \notag \\
&\simeq & \frac{1}{2}[1+ q_0^2 \pm \sqrt{(1+ q_0^2)^2-4 q_0^2\sin^2\theta} ~ ],
\eea
where $o(v_F^2/c^2)$ has been ignored. In the long-wave limit, {\it i.e.}, $c q\ll \omega_p$ or $q_0\ll 1$, \corr{we have two branches: an optical branch, the dispersion of which is $\omega_{o0}^2\equiv (\frac{\omega_o}{\omega_p})^2\simeq 1+q_0^2\cos^2\theta+q_0^4 \cos^2\theta\sin^2\theta$, and an acoustic branch, wherein $\omega_{a0}^2\equiv(\frac{\omega_a}{\omega_p})^2\simeq q_0^2\sin^2\theta-q_0^4 \cos^2\theta\sin^2\theta$.}

However $B_0$ can not be always ignored. To illustrate the conclusion, we consider the orders of $\omega_p$ and $\omega_c$. $\omega_c=\frac{e B_0}{m_g}=\corr{\frac{B_0 v_F^2}{E_F/e}}\sim 10^{13}\text{(rad/s)}$ (or $\hbar\omega_c\sim 10\text{meV}$) if $B_0\sim 1\text{T}$ and $E_F\sim 0.1\text{eV}$. $\omega_p =(\frac{e^2 n_0}{d \epsilon_0 m_g})^{1/2}=(\frac{e^3 E_F\text{(eV)}}{d \pi\epsilon_0 \hbar^2})^{1/2}\sim 3.7\times 10^{11}\text{(rad/s)}$ (or $\hbar\omega_p\sim 0.2\text{meV}$) if we also set $d\sim 1\text{cm}$. In other words, we can choose parameters to yield $\hbar\omega_p\ll\hbar\omega_c$. We can tune $\omega_p$ and $\omega_c$ by controlling $E_F$. For instance, when $E_F$ decreases, $\omega_p$ also decreases but $\omega_c$ increases. For the case $\omega_{c0}\neq 0$, another branch was observed, $\omega_h$, which was nominated as the high mode here, exhibiting an asymptotic behavior: $\omega_{h0}^2 \to 1+q_0^2$ when $\omega_{c0}\to 0$ and  $v_{F0}=0$.

\begin{center}
\begin{minipage}{0.85\textwidth}
\centering
\includegraphics[width=2.8in]{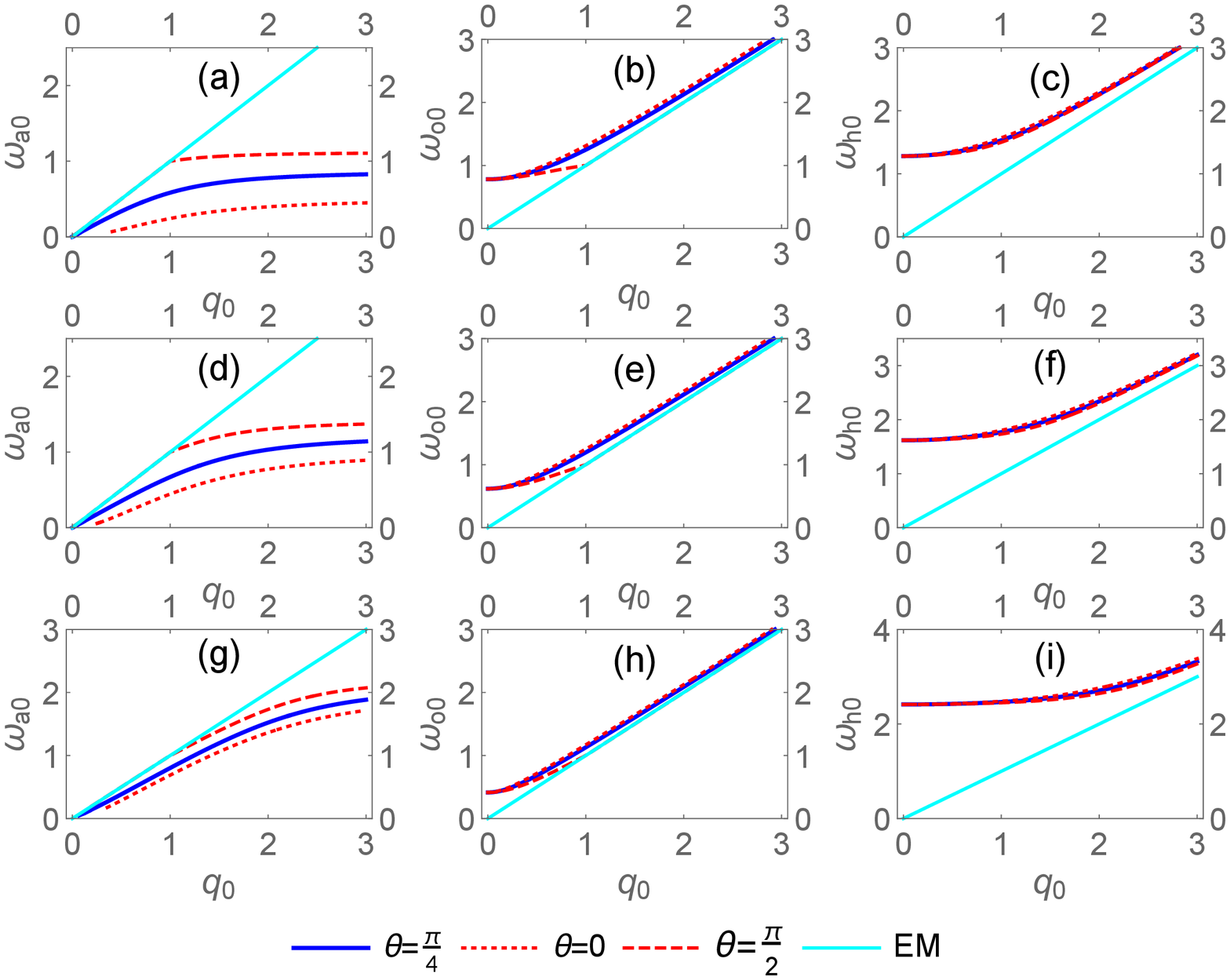}%
\figcaption{Dispersions of $\omega_a$(a,d,g), $\omega_o$(b,e,h), and $\omega_h$(c,f,i) at different angle. We set $\omega_{c0}$ to be 0.5(a,b,c),1.0(d,e,f), and 2.0(g,h,i) respectively. Here the symbol "EM" stands for the dispersion of electromagnetic radiation in the vacuum.
}
\end{minipage}
\setlength{\intextsep}{0.in plus 0in minus 0.1in} 
\end{center}

The discussions on the PP in such a structure are not sufficient, particularly when a background magnetic field exists. It is necessary to consider the influence of the background magnetic field on the PP. Neglecting $o(v_F^2/c^2)$ terms, we obtained dispersions of $\omega_a$, $\omega_o$, and $\omega_h$ at $\theta=0,\frac{\pi}{4}, \frac{\pi}{2}$ and $\omega_{c0}=0.5,1,2$ in Fig. 2, respectively.

From Fig. 2, we briefly list certain results based on the structure in the presence of the background magnetic field:
\begin{enumerate}
  \item When $\theta=\frac{\pi}{2}$, implying that the PP propagates along graphene layers, in the long-wave limit, $\omega_{a0}\simeq q_0$ but $\omega_{a0}< q_0$. In other words, in the long-wave limit, the dispersion of the acoustic branch is approximately equal to that of electromagnetic (EM) radiation in the medium, as highlighted in previous studies\ct{prime}. The statement holds in the presence of $B_0$. However, the stronger the background magnetic field, the narrower the scope of the statement, as shown in 2(a), 2(d), and 2(g).
  \item Generally, the larger the value of $\sin\theta$, the higher the frequency of the acoustic mode for the fixed $q_0$. In particular, in the absence of $B_0$, $\omega_{a0}\propto |\sin\theta|$, which was confirmed in \cite{ldq12}. However, as $B_0$ increases, the angular dependence becomes weaker, as shown in 2(a), 2(d), and 2(g).
  \item In addition to $\omega_a$, two other modes exist, $\omega_o$ and $\omega_h$, which have frequently been ignored \corr{by previous studies}. The modes are angular insensitive. The dispersion of $\omega_h$ is almost isotropic. Similarly, the angular dependence becomes weaker as $B_0$ increases.
  \item With an increase in $B_0$, $\omega_h$ increases, whereas $\omega_o$ decreases. In other words, the gap between the two dispersion curves becomes larger with an increasing $B_0$ because $\omega_o<\omega_h$ for certain $q_0$.
  \item The $\omega_a$ dispersion curve is below the EM radiation curve. $\omega_h$ and $\omega_o$ curves are above the EM curve for small values of $B_0$. If $B_0$ is sufficiently strong, {\it i.e.} $\omega_{c0}$ is adequately high, the $\omega_o$ and EM curves will intersect.
\end{enumerate}

To consider the coupling between the axion field and PP, we further introduce dimensionless quantities, $n_{10}=\frac{c^2}{\omega_p^2} n_1$, $d_0=\frac{\omega_p}{c} d$, $g_0=\frac{\sqrt{\hbar}\omega_p}{c}g$, $\phi_0=\frac{c}{\sqrt{\hbar}\omega_p}\phi$, \corr{$\omega_{\phi 0}=\omega_\phi/\omega_p$}, and $\omega_{g0}=\frac{m_g c^2}{\omega_p\hbar}$. \corr{ Note that when we say in the paper that $d$ is sufficient small we means that, from the above dimensionless quantities, $d_0<<1$, {\it i.e.}, $d\ll \frac{c}{\omega_p}$. In other word, if $\omega_0\sim 10^{11}\text{rad/s}$ was chosen, a sufficient small $d$ means that $d\ll 1\text{mm}$.}

Eqs. (\ref{res1}) can then be written as
\beea
[(q_0^2-\omega_0^2+\omega_{\phi 0}^2) +g_n^2 \frac{ q_0^2\cos^2\theta-\omega_0^2}{q_0^2-\omega_0^2}]\phi_0+
\frac{\ci g_n q_0\sqrt{\pi}\cos\theta}{\sqrt{d_0\omega_{g0}}
 v_{F0}(q_0^2-\omega_0^2)} n_{10} &=& 0 \notag \\
\frac{ \ci g_n q_0^3 v_{F0} \sqrt{\omega_{g0}d_0}(1+q_0^2-\omega_0^2) \cos\theta \sin^2\theta}{\sqrt{\pi}[\omega_0^2(1+q_0^2-\omega_0^2)^2-(q_0^2-\omega_0^2)^2\omega_{c0}^2]}\phi_0 +
[1-\frac{q_0^2(1+q_0^2-\omega_0^2)(2+v_{F0}^2(q_0^2-\omega_0^2))\sin^2\theta} {2[\omega_0^2(1+q_0^2-\omega_0^2)^2-(q_0^2-\omega_0^2)^2\omega_{c0}^2 ]} ]n_{10} &=& 0
\eea
where $g_n=g_0 A_m$ with the enlargement factor $A_m= \frac{\omega_{c0}\omega_{g0}^{3/2}v_{F0}}{\sqrt{\pi d_0}}$.

Rescaling $n_{10}$, we can rewrite the above equation as:
\bee
\left(
  \begin{array}{cc}
    (q_0^2-\omega_0^2+\omega_{\phi 0}^2) +g_n^2 \frac{ q_0^2\cos^2\theta-\omega_0^2}{q_0^2-\omega_0^2} & \ci\frac{g_n q_0^2
    \cos\theta\sin\theta}{q_0^2-\omega_0^2}\sqrt{\frac{(q_0^2-\omega_0^2)(1+q_0^2-\omega_0^2)}{\omega_{c0}^2(q_0^2-\omega_0^2)^2-\omega_0^2(1+q_0^2-\omega_0^2)^2}} \\
     -\ci\frac{g_n q_0^2
    \cos\theta\sin\theta}{q_0^2-\omega_0^2}\sqrt{\frac{(q_0^2-\omega_0^2)(1+q_0^2-\omega_0^2)}{\omega_{c0}^2(q_0^2-\omega_0^2)^2-\omega_0^2(1+q_0^2-\omega_0^2)^2}}  & 1+\frac{q_0^2(1+q_0^2-\omega_0^2)\sin^2\theta} {(q_0^2-\omega_0^2)^2\omega_{c0}^2 -\omega_0^2(1+q_0^2-\omega_0^2)^2)} \\
  \end{array}
\right)\phi_n^T
\equiv H \phi_n^T =0,
\ee
where $\phi_n=(\phi_0,n_{10})$, and $H$ is a $2\times 2$ Hermitian matrix. We ignored the term $v_{F0}^2(q_0^2-\omega_0^2)$ because $|v_{F0}^2(q_0^2-\omega_0^2)|\ll 1$.

The above equation shows that the "{\it effective}" coupling between the axion and PP in the system is not $g$ but $g_n$, meaning that the hybridization between the axion and PP, and therefore, their spectra are mainly determined by $g_n$ (and $\omega_{c 0}$, $\omega_{\phi 0}$). This is an important result. Under ordinary conditions, $g$ is considerably small; however, considering the obtained results, one can tune the parameters to enhance the coupling $g_n$. To determine how to enhance the coupling, we wrote the enlargement factor as $A_m=e\pi c(\frac{\epsilon_0\hbar c}{e^2})^{3/2}\frac{B_0 d}{E_F}\equiv 3.39\times 10^8\times\frac{B_0(T)d(cm)}{E_F(eV)}$. Therefore, one can increase $A_m$ by increasing the external background magnetic field and layer distances (it is important to note \corr{ the requirement that $d$ should be far smaller than the wave lengthof the hybridized state, as pointed out in the following, or roughly speaking, $d\ll \frac{c}{\omega_p}$}) or decreasing carrier concentration (or equivalently, gate voltage).

As expected, in the decoupled case, {\it i.e.} $g_n=0$, there are four modes: one corresponds to the axion and the other three correspond to the PP modes. In the coupled case, the axion mode hybridizes with three PP modes. If $\omega_{\phi 0}\geqq 1$, the axion mode  mainly hybridizes with the high and optical mode. In this study, we focus on the ultralight axion, $\omega_{\phi 0}\ll 1$. In this case the axion mainly hybridizes with the acoustic mode, particularly in the long wave limit.

We nominate the coupled axion/the acoustic mode as the axion hybridization state/acoustic hybridization state (AXHS/ACHS), as AXHS/ACHS also contains PP/axion constitution owing to the hybridization and AXHS/ACHS are pure axion/acoustic PP states when $g_n\to 0$.

The results of AXHS and ACHS are shown in Fig. 3 (a), where $\omega_{\phi 0}=0.1$, $\omega_{c0}=0.5$ and $\theta=\pi/4$. For the decoupled case, $g_n=0$, the AXHS dispersion is $\omega_{ax 0}=\sqrt{\omega_{\phi 0}^2+q_0^2}$ and the ACHS dispersion is the same as in Fig. 2, {\it i.e.}, $\omega_{a 0}\simeq q_0\sin\theta$ in the long wave limit. However, in the coupled case, for instance, $g_n=0.3$, the hybridization between the axion and acoustic PP occurs. The AXHS increases and the ACHS decreases owing to the hybridization. Therefore, the gap between them becomes larger. Additionally, in the long-wave limit, the ratio between $\omega_{a0}$ and $q_0$, $r(g_n)\equiv \lim\limits_{q_0 \to 0}\frac{\omega_{a0}}{q_0}$, should be smaller than $\sin\theta$, as shown in Fig. 3(a). \corr{To verify the result, we also list $r(g_n)$ at different angles for the cases $g_n=0, 0.03, 0.1, 0.3$ and $\omega_{\phi 0}=0.1$ in Fig. 3(b) and for the cases $g_n=0,0.005,0.01,0.05$ and $\omega_{\phi 0}=0.01$ in Fig. 3(c) respectively, which show that as $g_n$ deviates from zero, the ratio deviates from the uncoupled case $r(g_n=0)=\sin\theta$. Furthermore, the smaller $\omega_\phi$ is, the more significant the deviation. Fig. 3(c) exemplifies that the deviation is significant even when $g_n=0.005$ at the case $\omega_{\phi 0}=0.01$.
In the presence of hybridization, both ACHS and AXHS spectra can be detected using common photoelectric methods in the laboratory. Therefore, it is a suitable method to measure the deviation of $r(g_n)$ to judge the presence or absence of axions in the universe. As the measurement is independent of resonance, the detectable axion mass range is broad. In particular, the measurement becomes more sensitive as the axion mass decreases. From Fig. 3(c) we conclude that when $g_n=0.005$ was chosen, the axion is detectable as long as $\omega_{\phi 0}<0.01$. In other word, the detection range for axion is roughly $\omega_{\phi 0}<0.01$. As a "back of the envelope" estimation, if one chooses that $\omega_p\sim 10^{11}\text{rad/s}$ or $\hbar\omega_p\sim 0.1\text{meV}$, detectable range of axion mass is $m_\phi c^2\stackrel{<}{\sim} 1\mu \text{eV}$.}

\begin{center}
\begin{minipage}{0.85\textwidth}
\centering
\includegraphics[width=2.8in]{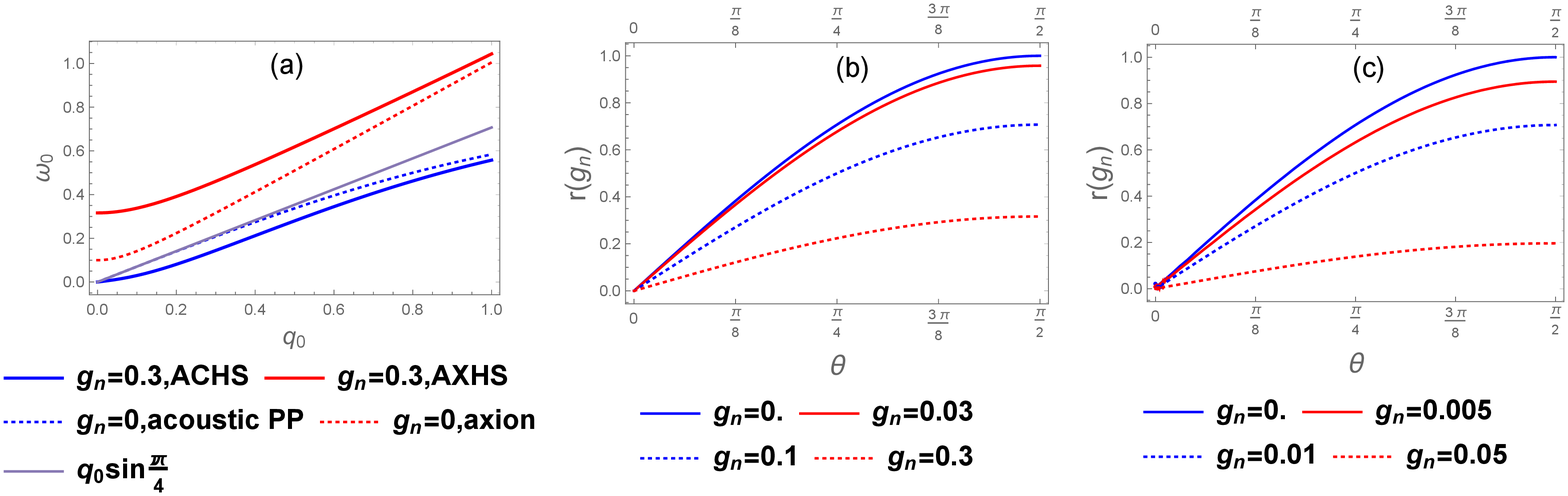}%
\figcaption{Dispersion relations of AXHS and ACHS (a) and ratios between $\omega_{a0}$ and $q_0$ at different angle in the long wave limit.  Besides $\omega_{c0}=0.5$, we also set
$\theta=\frac{\pi}{4}$(a), $\omega_{\phi 0}=0.1$ (a,b), and $\omega_{\phi 0}=0.01$(c). In (b,c) ratio at the decoupled case is just $r(g_n=0)=\sin\theta$.
}
\end{minipage}
\setlength{\intextsep}{0.in plus 0in minus 0.1in} 
\end{center}

Furthermore, it is important to study the influence of the background magnetic field on the AXHS spectrum or ACHS spectrum, which was shown in Fig. 4. The AXHS spectrum is insignificantly affected by $\omega_{c0}$ and the ACHS spectrum is mildly affected when $g_n$ remains constant. The results revealed that the impact of background magnetic field on the AXHS spectrum is mostly due to $g_n$ and not $\omega_{c0}$; meanwhile, both $\omega_{c0}$ and $g_n$ influence the ACHS spectrum.

\begin{center}
\begin{minipage}{0.85\textwidth}
\centering
\includegraphics[width=2.8in]{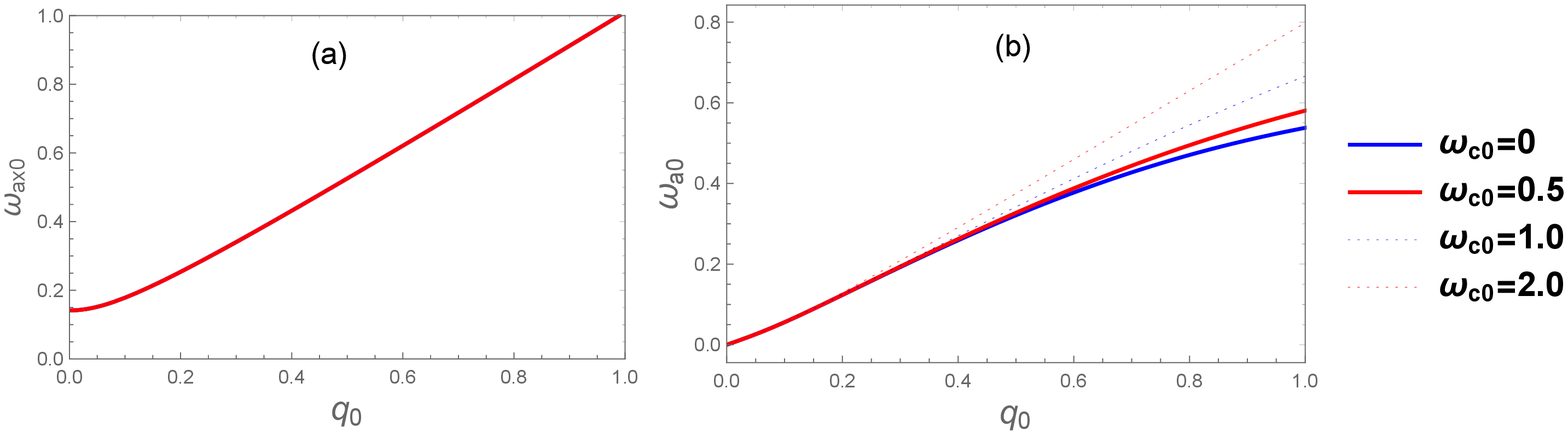}%
\figcaption{Dispersion relations of AXHS (a) and ACHS (b) at different background magnetic field ($\omega_{c0}$). Here we set $\omega_{\phi 0}=0.1$, $\theta=\frac{\pi}{4}$ and $g_n=0.1$.
}
\end{minipage}
\setlength{\intextsep}{0.in plus 0in minus 0.1in} 
\end{center}

\section{Conclusions}
We investigated and discussed the coupling between the axion and PP as well as the spectrum of AXHS and ACHS in a doped graphene periodic structure. The discussion highlighted that the effective coupling between the axion and PP, $g_n$, depends on parameters such as carrier concentration, interlayer distance, and background magnetic field. Accordingly, these parameters can be tuned to enhance effective coupling.

We further studied the hybridization between the axion and PP, particularly, acoustic PP stemming from the effective coupling $g_n$. The hybridization modified the dispersion relations of AXHS and ACHS efficiently. An observable, $r(g_n)$, was proposed to detect axions accordingly. As long as the axion mass is sufficiently small, the deviation of $r(g_n)$ from uncoupled behavior $\sin\theta$ is significant even when the effective coupling is not large.

With the use of the above two developments, we believe it is possible to detect axions in the laboratory. Furthermore, as the measurement is independent on the resonance phenomenon, the detectable range of axion mass is wide. Our proposed system provides an easy-to-realize and sensitive platform for detecting axions in the universe.








\end{document}